\documentclass[aps, pra, twocolumn, superscriptaddress, amsmath, 
tightenlines, longbibliography]{revtex4-2}

\usepackage{amssymb}
\usepackage{amsmath}
\usepackage{cases}
\usepackage{dcolumn}
\usepackage{graphicx}
\usepackage{mathrsfs}
\usepackage{appendix}
\usepackage{graphicx}
\usepackage{booktabs}
\usepackage{colortbl}
\usepackage{float}

\definecolor{Dred}{RGB}{190,0,0}

\usepackage{url}
\usepackage[colorlinks]{hyperref}
\hypersetup{%
	plainpages=true,
	breaklinks=true,       
	hypertexnames=false,  
	pageanchor=true,
	colorlinks=true,
	linkcolor={blue},
	citecolor={red},
	urlcolor={blue},
	anchorcolor={black}
}

\def \hide#1{}

\begin{document}
\title{Harnessing spontaneous emission of correlated photon pairs \\ from 
ladder-type giant 
atoms}

\author{Zhao-Min Gao}
\affiliation{Institute of Theoretical Physics, School of Physics, Xi'an Jiaotong University, Xi'an 710049, People’s Republic of China}

\author{Jia-Qi Li}
\affiliation{Institute of Theoretical Physics, School of Physics, Xi'an Jiaotong University, Xi'an 710049, People’s Republic of China}

\author{Ying-Huan Wu}
\affiliation{Institute of Theoretical Physics, School of Physics, Xi'an Jiaotong University, Xi'an 710049, People’s Republic of China}

\author{Wen-Xiao Liu}
\affiliation{Institute of Theoretical Physics, School of Physics, Xi'an Jiaotong University, Xi'an 710049, People’s Republic of China}
\affiliation{Department of Electronic Engineering, North China University of Water Resources and Electric Power, Zhengzhou 450046, People’s Republic of China}

\author{Xin Wang}
\email{wangxin.phy@xjtu.edu.cn}
\affiliation{Institute of Theoretical Physics, School of Physics, Xi'an Jiaotong University, Xi'an 710049, People’s Republic of China}

\date{\today}

\begin{abstract} 
The realization of correlated multi-photon processes usually depends on the interaction between nonlinear media and atoms. However, the nonlinearity of optical materials is generally weak, making it still very challenging to achieve correlated multi-photon dynamics at the few-photon level. Meanwhile, giant atoms, with their capability for multi-point coupling, which is a novel paradigm in quantum optics, mostly focus on the single photon field. In this work, using the method described in \href{https://link.aps.org/doi/10.1103/PhysRevResearch.6.013279}{Phys. Rev. Res. 6. 013279 (2024)}, we reveal that the ladder-type three-level giant atom spontaneously emits strongly correlated photon pairs with high efficiency by designing and optimizing the target function. In addition, by encoding local phases into the optimal coupling sequence, directional two-photon correlated transfer can be achieved. This method does not require a nonlinear waveguide and can be realized in the conventional environment. We show that the photon pairs emitted in both the bidirectional and the chiral case exhibit strong correlation properties in both time and space. Such correlated photon pairs have great potential applications for quantum information processing. For example, numerical results show that our proposal can realize the two-photon mediated cascaded quantum system.
\end{abstract}
\maketitle
\section{Introduction}
Giant atoms, which are different from point-like small atoms, can be comparable to or much larger than the wavelength of the interacting field in size \cite{PhysRevA.108.043709,Xu_2024,PhysRevA.106.063703,PhysRevA.107.013710,PhysRevA.108.023728,Xinyu_2023,PhysRevA.108.053718}. Therefore, the dipole approximation becomes invalid, enabling giant atoms to achieve multi-point coupling with waveguides \cite{PhysRevLett.128.223602,Longhi:20,PhysRevA.103.023710,PhysRevResearch.4.023198,PhysRevA.90.013837,PhysRevA.101.053855,PhysRevA.108.013704}. At present, most studies on giant atoms focus on the field of single-photon processes, such as frequency-dependent couplings \cite{Dulei_2023,PhysRevA.107.063703,PhysRevLett.120.140404,PhysRevA.107.023716,jia2023atom,PhysRevA.105.023712} and chiral quantum optics \cite{PhysRevA.105.023712,PhysRevA.104.023712,PhysRevA.107.023705,PhysRevResearch.2.043184,Xiao_2022,PhysRevLett.126.043602,PhysRevResearch.2.043014,PhysRevA.102.033706,Chen_2022}. These studies have led to significant advancements in controlling light-matter interactions at the single-photon level. However, only a few studies have delved into the multi-photon dynamics of giant atoms \cite{wang2024nonlinear,PhysRevA.109.023720}, for which nonlinear media are required.
 
Meanwhile, multi-photon dynamics are crucial for building long-range, high-fidelity quantum information and quantum communication networks \cite{PhysRevLett.132.047001,DELLANNO200653,Zhang_2019,Liu_2022,wang2024longrange,Metcalf_2013,PhysRevLett.121.143601,PhysRevLett.123.253601}, typically implemented in nonlinear media \cite{LiPeng_2020,Adarsh_2020,PhysRevResearch.4.023026,PhysRevResearch.4.023026,Cantu_2020,HE2019155,HEGUANG_2014,Murti2021,Thibault_2022,PhysRevLett.126.083605,PhysRevLett.124.213601}. Multi-photon processes have made significant progress in the fields of quantum entanglement dynamics, quantum interferometry, dissipative quantum dynamics and optical parameter processes \textit{et al.} \cite{chuk_2023,PhysRevLett.125.163602,RevModPhys.84.777,Barbieri_2009,Llewellyn_2019,PhysRevApplied.19.L011002,PhysRevLett.110.113601,PhysRevB.107.235304,PhysRevApplied.20.044080,PhysRevApplied.19.L011002,PhysRevA.107.012607,RevModPhys.73.565}. However, there are still great challenges in the field of multi-photon problems, such as photon loss, nonlinear effects and quantum noise, which limit the stable transmission of multi-photon states \cite{Qian_2023,PhysRevA.108.013705,PhysRevA.106.023708,Jeannic_2022,Wanglei_2022,PhysRevA.92.053834,PhysRevLett.132.203602,Fraser_2021}. Moreover, the weak nonlinearity of the optical materials imposes great limitations on the multi-photon dynamics.

In recent years, the exploration of quantum optics within the structured environment of giant atoms has been proposed \cite{PhysRevResearch.6.013279}. Unlike previous studies of giant atoms which only considered a few coupling points, this study proposes to use the coupling of multiple points $x_1,...,x_N$ with interaction strength $g(x_i)$, where the distances between different coupling points are unequal. The optimal coupling sequence is obtained by using the sequential quadratic programming algorithm to optimize the  objective function of the interaction between the giant atom and the conventional waveguide in momentum space, and by using an inverse Fourier transform to obtain the coupling sequence in real space. With this method, exotic quantum phenomena such as simulated band-gap environment and broadband chiral emission are realized. Here, we extend this optimization method to multi-level giant atom. We find that the multi-photon processes which are correlated in both time and space can be realized in conventional linear waveguides without requiring any nonlinear media. Moreover, we can achieve high-fidelity quantum state transfer between two three-level giant atoms under this scheme.
 
In this paper, we propose to use a ladder-type three-level giant atomic platform to achieve correlated two-photon processes in a linear waveguide. Firstly, we consider the bidirectional propagation of photons. By calculating the spontaneous decay process numerically and analytically, we prove that the three-level giant atom emits two correlated photons simultaneously, rather than in cascades. In addition, we obtain the optimal coupling sequence by designing the objective function and optimizing it. Using this coupling sequence, we explore the spatial correlation properties of the two photons. Secondly, we also focus on the case where the correlated two photons can only propagate in the same direction. Finally, the two-photon mediated cascaded quantum system based on our method is proposed.
\begin{figure*}[ht!]
	\centering\includegraphics[width=17cm]{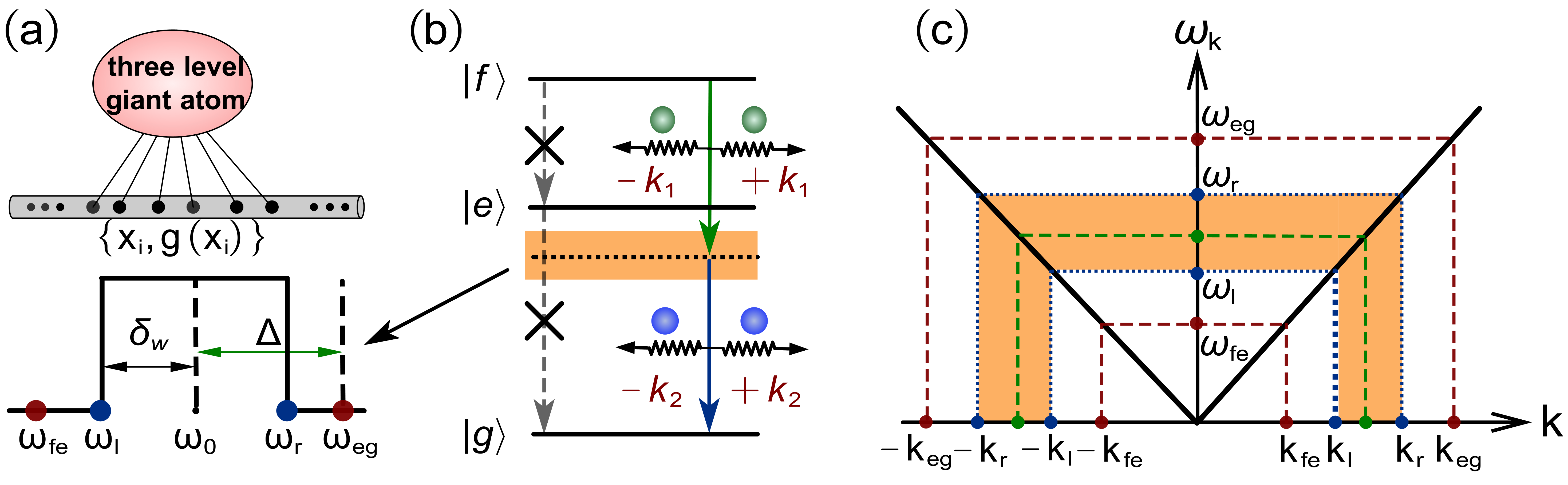}
	\caption{(a) Sketch of the bidirectional model: A three-level giant atom couples to the linear conventional wavguide at positions $x_1$, ..., $x_N$, which photons can travel in both directions. The diagram below is a sketch of the coupling function in the form of a window function. (b) Energy level diagram of the ladder-type three-level giant atom. (c) The dispersion relation of the linear waveguide, $\omega_k=c|k|$. The orange region represents the frequency range at which the giant atom can be coupled to the waveguide.}
	\label{fig1}
\end{figure*}

\section{Bidirectional two-photon emission}
\subsection{Model and Equations}
\label{section1_a}
As shown in Fig.~\ref{fig1}(a), we consider a ladder-type three-level giant atom with ground state $|g\rangle $ and first (second) excited state $|e\rangle $ ($|f\rangle $), coupled to a one-dimensional conventional waveguide. The three-level giant artificial atom can be easily implemented with superconducting qubit circuits \cite{PhysRevA.90.013837,RevModPhys.93.025005,PhysRevA.76.042319}.  We define the energy between the $ i$th and $j$th energy levels as $\omega_{ij}$, where $i,j\in \left\{ f,e,g \right\} $ The Hamiltonian of the system is (setting $\hbar$=1)
\begin{eqnarray}
	H&=&\omega _{fg}\sigma _{fg}^{+}\sigma _{fg}^{-}+\omega _{eg}\sigma _{eg}^{+}\sigma _{eg}^{-} +\sum_k{\omega _ka_{k}^{\dagger}a_k}\notag \\
	&+&\sum_k{g_ka_{k}^{\dagger}\sigma _{fe}^{-}} 
	+\sum_{k'}{g_{k'}a_{k'}^{\dagger}\sigma _{eg}^{-}}+\mathrm{H}.\mathrm{c}.,
\end{eqnarray}
where $a_k $ ($a_k^\dagger$) is photon annihilation (creation) operator, with $k$ being the wave vector in momentum space, and $\sigma _{ij}^{+}=|i\rangle \langle j| $ denotes the atomic transition operator, with $i,j\in \left\{ f,e,g \right\} $. We consider the giant atom coupling to the waveguide at multiple points $x_1,...,x_N$ with distinct interaction strength $g(x_i)$. The interaction strength in momentum space is expressed as $g_k=\sum_{x_i}{g\left( x_i \right) e^{-ikx_i}}$ \cite{PhysRevA.76.042319,RevModPhys.93.025005,Wang_2022}. In our model, the waveguide is considered to be conventional without periodical structure, with linear dispersion $\omega _k=c\left| k \right|$, with $c$ being the group velocity. 

In the rotating-wave approximation, the total Hamiltonian of the system is 
\begin{eqnarray}
	H_I&=&\sum_k{\left( \omega _k-\omega _{eg} \right) a_{k}^{\dagger}a_k} \notag \\
	&+&\sum_k{g_ka_{k}^{\dagger}e^{2i\Delta t}\sigma _{fe}^{-}} +\sum_{k'}{g_{k'}a_{k'}^{\dagger}\sigma _{eg}^{-}}+\mathrm{H}.\mathrm{c}.,
\end{eqnarray}
where $\Delta=\omega _{eg}-\omega_0=\omega_0-\omega _{fe}$ [see Fig.~\ref{fig1}(b)]. In the two-excitation subspace, the state of the system can be expressed as
\begin{eqnarray}
	\begin{split}
		|\psi \left( t \right) \rangle &=c_f\left( t \right) |f,0\rangle +\sum_k{c_{e,k}\left( t \right)|e,k\rangle}\\
		&+\sum_{k,k\prime}{c_{g,k,k\prime}\left( t \right) |g,k,k'\rangle}  .
	\end{split}
\end{eqnarray}
Here, $c_{g,k,k'}(t)$ denotes the probability amplitude of the giant atom being in the ground state, while simultaneously exciting two photons with wave vector $k$ and $k' $ in the waveguide. Additionally, $c_{e,k}(t)$ ($c_f(t)$) is the probability amplitude of the first (second) excited state.

For small atoms, the coupling to the waveguide is effectively localized at a single point $\delta(r)$ in real space, leading to a constant coupling across all momentum modes, i.e.,  $\int{\delta \left( r \right) e^{-ikr}dr}=1 $. Therefore, achieving correlated two-photon radiation is challenging, and only cascaded emission can typically be observed in small atom systems. In contrast, the unique properties of giant atoms allow for multi-point coupling with the waveguide. The coupling strength of each mode in k-space is no longer identical, which is an important feature of giant atoms. This distinctive feature enables us to design the coupling sequence, thereby suppressing single-photon radiation and facilitating correlated two-photon emission. Additionally, giant atoms are typically implemented in superconducting circuits, where strong coupling with the waveguide can be readily achieved. This strong coupling effectively compensates for the effects of large detuning on the two-photon emission rate. Furthermore, the high coupling efficiency in superconducting circuits allows us to disregard the impact of decoherence on the system's dynamics \cite{Mohammad_2019,PhysRevB.100.035311}. Therefore, we propose employing a three-level giant atom to realize two-photon processes.

Initially, the three-level atom is considered to be in the second excited state, while the waveguide is in its vacuum state, i.e., $|\psi \left( t=0 \right) \rangle =|f,0\rangle $. According to Schr$\ddot{\mathrm{o}}$dinger equation, we derive the following coupling equations
\begin{subequations}
	\begin{numcases}{}
		i\dot{c}_f\left( t \right) =\sum_k{g_{k}^{\ast}e^{-2i\Delta t}c_{e,k}\left( t \right)},\\
		\label{c_ft}
		i\dot{c}_{e,k}\left( t \right) =\left( \omega _k-\omega _{eg} \right) c_{e,k}\left( t \right) \notag\\+g_ke^{2i\Delta t}c_f\left( t \right) +\sum_{k'}{g_{k'}^{\ast}c_{g,k,k'}\left( t \right)},\\
		i\dot{c}_{g,k,k'}\left( t \right) =\left( \omega _k+\omega _{k'}-2\omega _{eg} \right) c_{g,k,k'}\left( t \right) \notag \\
		+g_{k'}c_{e,k}\left( t \right).
		\label{c_gt}
	\end{numcases}
\end{subequations}

\begin{figure*}
	\centering\includegraphics[width=17cm]{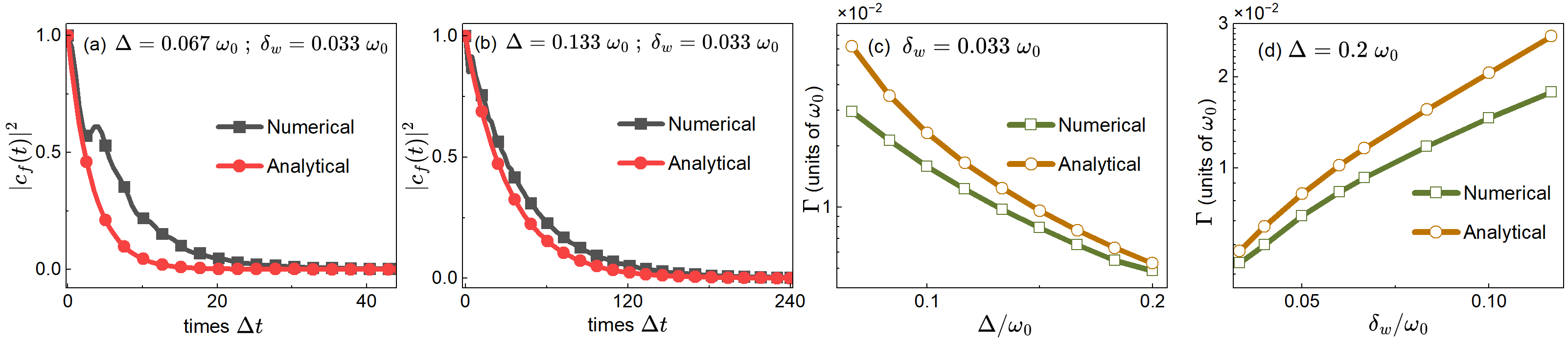}
	\caption{(a), (b) The numerical (see Appendix.~\ref{appendix1}) and analytical results (c.f. Eq.~\ref{gamma}) of the spontaneous 
	emission when the objective function takes the form of window function, 
	setting $\delta_w/\omega_0=0.033$ with $\Delta/\omega_0=$ (a) 0.067, (b) 0.133. (c) The spontaneous decay rate changes with $\Delta$, with $\delta_w/\omega_0=0.033$. (d) The spontaneous decay rate changes with the width of the window function, 
	with $\Delta/\omega_0=0.2$.}
	\label{fig2}
\end{figure*}

The spontaneous decay process of the three-level giant atom we aim to achieve is correlated emission where two photons are emitted simultaneously, rather than cascaded decay where the two photons are emitted sequentially . To achieve two-photon correlated emission for a three-level giant atom, the cascaded decay process within the atom must be suppressed. We assume that the giant atom decouples from the waveguide near the two transition frequencies $\omega_{fe} $ and $\omega_{eg}$, but is significantly stronger within an intermediate range around $\omega_0=\omega_{fg}/2$. Under these conditions, there is no mode in the waveguide resonating with the two transition frequencies of the giant atom, suppressing single-photon processes. On the contrary, the photons will be significantly excited via a second-order process around $\omega_0$. The three-level emitter is unable to remain in the first excited state but may instead transit to an intermediate state. However, because this intermediate state is not a natural energy level, the atom is impossible to maintain energy conservation in this state and will rapidly decay to the ground state. These two transitions, from the second excited state to the intermediate state, and then from the intermediate state to the ground state, may occur nearly simultaneously. For simplicity in calculations, we assume that the objective function in momentum space following the form of window function, specifically confined to the range $[\pm k_l,\pm k_r]$, with $ c|k_0|=\omega_{0}$ and $k_{r(l)}=k_0\pm \delta_w/c$, and a window function width of $2\delta_w/c$. In this work, we take $k_0$ and $\omega_0$ as units of wave vector and frequency, respectively.

Therefore, we calculate the evolution by using adiabatic elimination, i.e., $\dot{c}_{e,k}\left( t \right) =0$. We obtain
\begin{gather}
	c_{e,k}\left( t \right) =-\frac{g_ke^{2i\Delta t}c_f\left( t \right) +\sum_{k'}{g_{k'}^{\ast}c_{g,k,k'}\left( t \right)}}{\omega _k-\omega _{eg}}.
	\label{c_e,k}
\end{gather}
After substituting Eq.~(\ref{c_e,k}) into Eq.~(\ref{c_ft},\ref{c_gt}), we can calculate $c_{g,k,k'}(t)$ as
\begin{gather}
	c_{g,k,k'}\left( t \right) =i\frac{g_{k'}g_k}{\omega _k-\omega _{eg}}\int_0^t{e^{-i\delta' (t-t')}e^{2i\Delta t}c_f\left( t' \right) dt'},
	\label{gkk}
\end{gather}
with
\begin{gather}
	\delta'= \left( \omega _k+\omega _{k'}-2\omega _{eg} \right) -\frac{g_{k'}^{2}}{\omega _k-\omega _{eg}}.
\end{gather}

The Stark shift of the state $|f,0\rangle $ is derived as
\begin{gather}
\delta\omega_f=\sum_k{\frac{g_{k}^{2}}{\omega _k-\omega _{eg}}}.
\end{gather}
 Therefore, the probability amplitude $c_f(t)$ is rewritten as
\begin{equation}
	\begin{split}
		\dot{c}_f\left( t \right) &=i\delta\omega_fc_f(t)  \\
		&-\sum_{k,k'}{\frac{g_{k}^{2}g_{k'}^{2}}{\left( \omega _k-\omega _{eg} \right) ^2}\int_0^t{e^{-i(\delta'+2\Delta) t}c_f\left( t' \right) dt'}}.
		\label{c_f1}
	\end{split}
\end{equation}
By defining 
\begin{gather}
	\tilde{c}_f\left( t \right) = c_f\left( t \right) e^{-i\sum_k{\frac{g_{k}^{2}}{\omega _k-\omega _{eg}}}t},
\end{gather}
we can renormalize energy level of $|f\rangle $ with dynamical Stark shift as
\begin{gather}
	\dot{\tilde{c}}_f\left( t \right) =-\sum_{k,k'}{\frac{g_{k}^{2}g_{k'}^{2}}{\left( \omega _k-\omega _{eg} \right) ^2}\int_0^t{e^{-i\delta_{k,k'} \left( t-t' \right)}\tilde{c}_f\left( t' \right) dt'}},
	\label{c_dot}
\end{gather}
where $\delta_{k,k'}=\omega_k+\omega_{k'}-\omega_{fg}$ is the two-photon 
detuning. The modes satisfying $\omega_k+\omega_{k'}=\omega_{fg}$ have great 
contribution to the dynamics of the two-photon process (cf. 
Eq.~(\ref{c_dot})). By employing the Wigner-Weisskopf theory 
\cite{scully_zubairy_1997}, the summation over $k$ is transformed into an 
integral, i.e., $\sum{_k\rightarrow 
	\frac{L}{2\pi}\int_{-\pi}^{\pi}{dk}}$, with $L\rightarrow\infty$ being 
	the waveguide length, and Eq.~(\ref{c_dot}) is written as
\begin{gather}
	\dot{\tilde{c}}_f\left( t \right) =-\left( \frac{L}{2\pi} \right) 
	^2\int_0^t 
	\int_{-\infty}^{\infty}{\int_{-\infty}^{\infty}{\frac{g_{k}^{2}g_{k'}^{2}}{\left(
	\omega _k-\omega _{eg} \right) ^2}}}  \notag \\
	e^{-i\delta_{k,k'} \left( t-t' \right)}\tilde{c}_f\left( t' \right) dkdk'dt'.
\end{gather}
 Given that we consider a linear waveguide, where $\omega_k=c\left| k \right|$, and obtain
\begin{gather}
	\dot{\tilde{c}}_f\left( t \right) =-\left( \frac{L}{2\pi} \right) 
	^2\frac{1}{c^2}\int_0^t\int_{-\infty}^{\infty}{\int_{-\infty}^{\infty}{\frac{g_{k}^{2}g_{k'}^{2}}{\left(
	 \omega _k-\omega _{eg} \right) ^2}}} \notag \\
	e^{-i\delta_{k,k'} \left( t-t' \right)}\tilde{c}_f\left( t' \right) d\omega _kd\omega _{k'}dt'.
	\label{c_f2}
\end{gather}
To calculate the spontaneous decay rate, we perform a variable substitution for $\omega_k$ and $\omega_{k'}$, i.e.,
\begin{gather}
	\begin{cases}
		\omega_{+}=\omega _k+\omega _{k'},\\
		\omega_{-}=\omega _k-\omega _{k'}.\\
	\end{cases}
\end{gather}
We calculate the frequency-dependent component in Eq.~(\ref{c_f2}) and obtain
\begin{gather}
	\int_{0}^{\infty}{\int_{0}^{\infty}{\frac{1}{\left( \omega 
	_k-\omega 
	_{eg} \right) ^2}e^{-i\delta_{k,k'} \left( t-t' \right)}d\omega _kd\omega 
	_{k'}}} \notag
	\\
	=2\int_{0}^{\infty}{-\frac{1}{\omega_{+}+\omega_{-}-2\omega 
	_{eg}}\mid_{-2\delta_w}^{2\delta_w}e^{-i\left( \omega_{+}-\omega _{fg} 
	\right) \left( t-t' \right)}d\omega_{+}}.
\end{gather}

 We assume that the coupling strength is much smaller than the bandwidth 
 around $\omega_0$. 
 Moreover, since the atomic spontaneous radiation is centered at the 
 transition frequency, we can employ the integral 
 $$\int_{0}^{\infty}{e^{-i\left( \omega_{+}-\omega _{fg} \right) \left( 
 t-t' \right)}d\omega_+}\simeq 2\pi \delta \left( t-t' \right) ,$$ and within the frequency range $[\omega_l,\omega_r]$ of the window function, the coupling strength reaches its maximum at $g_k=g_{k_0}$. Therefore, we obtain
\begin{gather}
	\dot{\tilde{c}}_f\left( t \right) =\frac{4L^2|g_{k_0}|^4\delta _w}{\pi 
	c^2}\frac{\tilde{c}_f\left( t \right)}{\Delta ^2-\delta _{w}^{2}}  .
\end{gather}
Therefore,  the exponential decay is calculated as
\begin{gather}
c_f\left( t \right) =e^{-\frac{\Gamma}{2}t}, \qquad\Gamma =\frac{8L^2|g_{k_0}|^4}{\pi c^2}\frac{\delta _w}{\Delta ^2-\delta _{w}^{2}},
\label{gamma}
\end{gather}
with $\Gamma$ representing the spontaneous decay rate of emitting two correlated photons, note that this rate depends on the width of the window function and $\Delta$. As depicted in Fig.~\ref{fig2}(a,b), we numerically and analytically show the evolution of the three-level giant atom for different values $\delta_w$ and $\Delta$, observing discrepancies between numerical and analytical results. To determine appropriate values, we explore how the agreement between numerical and analytical results varies with changes in $\delta_w$ and $\Delta$ [see Fig.~\ref{fig2}(c,d)]. We find that the agreement improves as the window function narrows or as $\omega_{eg}$ moves further from $\omega_0$. To maintain a high two-photon emission rate, it is crucial to minimize the detuning between the first excited state and the band edge, specially keeping $\Delta-\delta_w<0.1\omega_0$. In the implementation using superconducting circuits, the strong coupling between the giant atoms and the waveguide helps to offset the effects of detuning, ensuring efficient emission. For our calculations, we set the parameters $\Delta=0.15\omega_0$ and $\delta_w=0.1\omega_0$ to optimize performance.

\subsection{Optimization for bidirectional case}
\label{sectionb}
\begin{figure}[h]
	\centering\includegraphics[width=8cm]{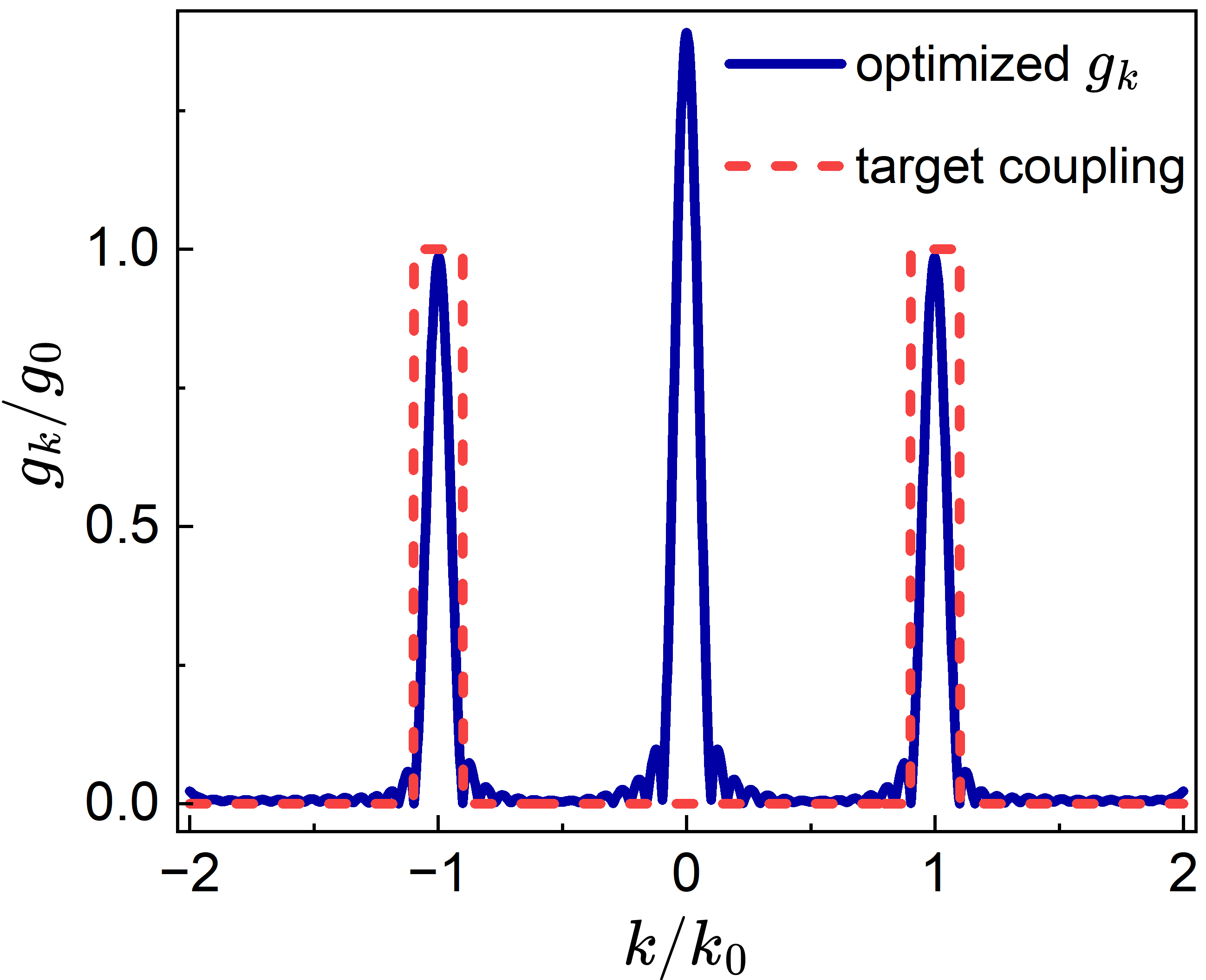}
\caption{The optimizing coupling function in momentum space for bidirectional 
correlated two-photon decay. The dashed red line represents the objective 
function, and the solid blue line is the optimized coupling function. The 
real-space coupling sequence $\left\{ x_i/x_0,g\left( x_i \right)/g_0 \right\} $ is 
listed in Table.~\ref{table1} [see Appendix.~\ref{appendix2}]. We set 
$|k_r|/k_0=1.1$ ($|k_l|/k_0=0.9$) and $N=50$.}
	\label{fig3}
\end{figure}

\begin{figure}[h]
	\centering\includegraphics[width=8cm]{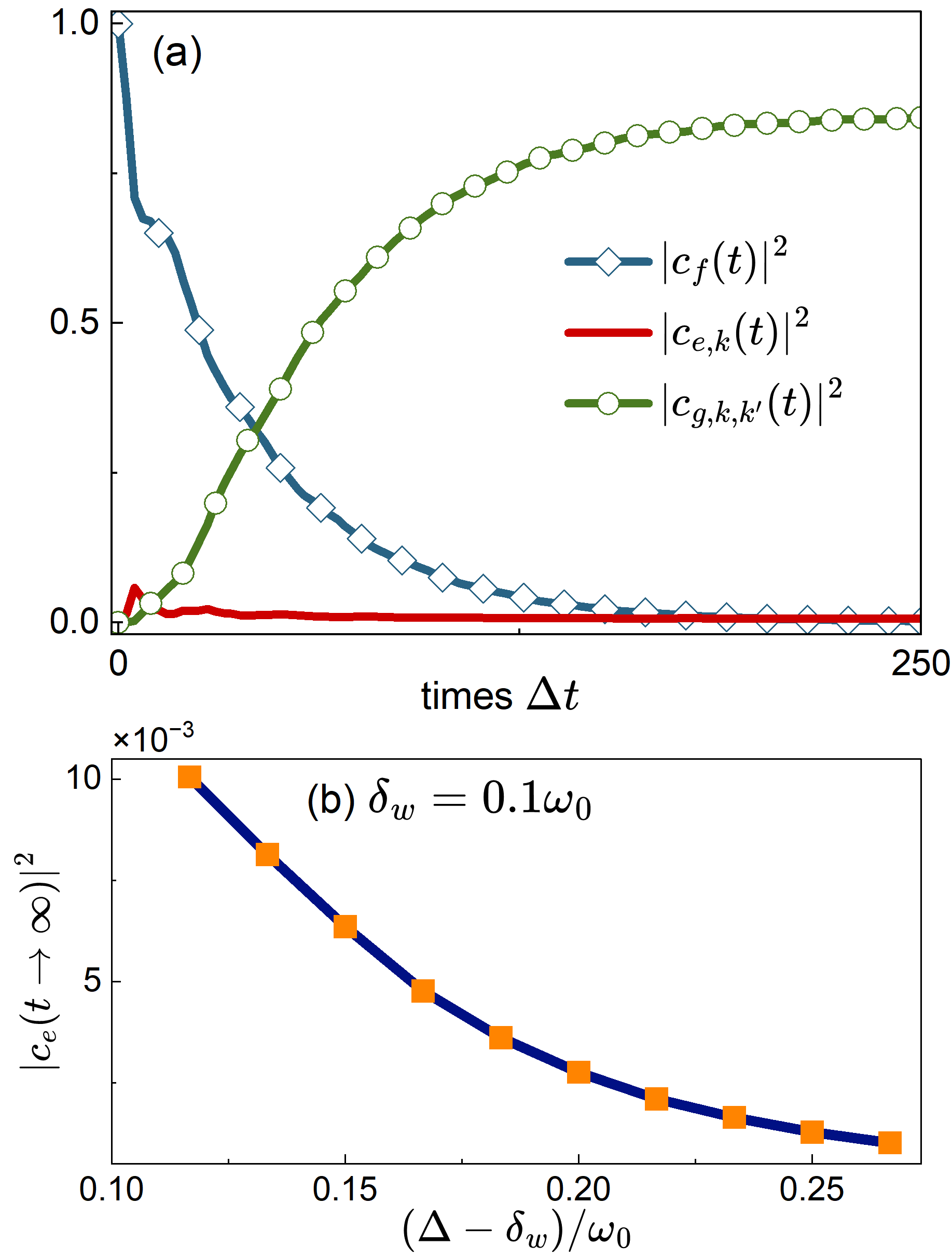}
	\caption{(a) The evolution of probability amplitudes of the three energy states $|f\rangle$, $|e\rangle$ and $|g\rangle$ of the giant atom with time. We set the parameter as $\omega_{eg}/\omega_0=1.15$ and $g_0=0.008$. (b) The probability of the giant atom's transition to the intermediate state varies with $\Delta$, with $\delta_w/\omega_0=0.1$.}
	\label{fig4}
\end{figure}

\begin{figure*}
	\centering\includegraphics[width=17cm]{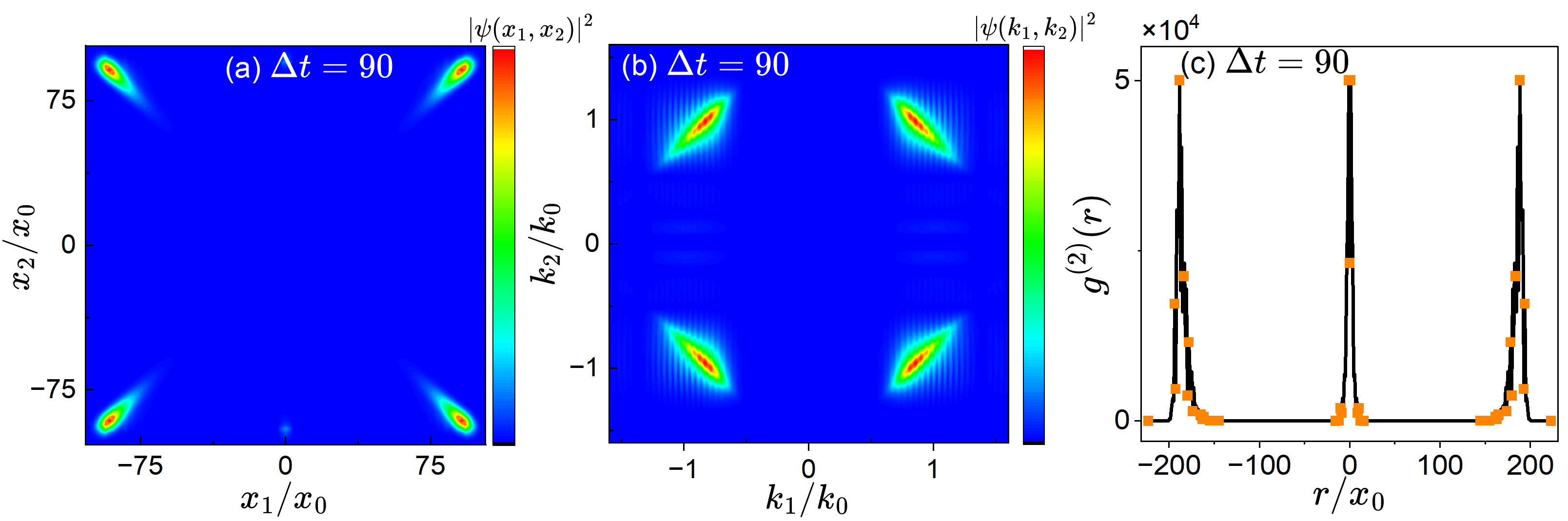}
	\caption{(a) The two-photon field distributions in real-space at $\Delta t=90$ . (b) The two-photon field distribution in momentum space. (c) The 
	second-order correlation function of two photons in the case of 
	bidirectional propagation. We set $g_0=0.02$ and $\omega_{eg}/\omega_0=1.15$.}
	\label{fig5}
\end{figure*}

In this section, we present the method to realize the optimization of target coupling for giant atoms. We assume that the coupling strength of the right $(k > 0)$ and left directions $(k < 0)$ are identical, i.e., $g_{-k} = g_k^\ast$, indicating that the photon propagates bidirectionally in the waveguide \cite{PhysRevResearch.6.013279}. By designing an objective function in the form of a window function in momentum space, the coupling strength is required to be exactly zero at the atomic transition frequencies between the three energy levels, as shown in Fig.~\ref{fig1}(b), and maximum at $\omega_0$, i.e.,
\begin{gather}
	g_{k}^{T}/g_0=\begin{cases}
		0,\quad \left\{ \omega _{fe}-\delta _w<c\left| k \right|<\omega _{fe}+\delta _w \right\} ,\\
		1,\quad \left\{ \omega _0-\delta_w <c\left| k \right|<\omega _0+\delta_w \right\} ,\\
		0,\quad \left\{ \omega _{eg}-\delta _w<c\left| k \right|<\omega _{eg}+\delta _w \right\} .\\
	\end{cases}
	\label{g_k}
\end{gather}
To obtain the coupling sequence in real space, we define the constraint condition as
\begin{gather}
	C_m=\int_{-k_{\max}}^{k_{\max}}{dk\left| \left| g_k \right|-\left| g_{k}^T \right| \right|w\left( k \right)},
\end{gather}
which represents the difference between the optimized coupling function $g_k$ and the objective coupling function $g_k^T$, with $w(k)$ being the weight function which describes the similarity between the optimized and objective functions in different regimes. In our work, to achieve correlated two-photon emission, we focus on the modes around $\pm k_0$ and $\pm k_{fe} $ $(\pm k_{eg})$. Due to the large detuning effect, the modes far away from these frequencies contribute little to the dynamics of the system. Therefore, we assign much greater weights to these three regions mentioned in Eq.~(\ref{g_k}) compared to others.

The coupling mechanism of linear quantum electrodynamics (QED) elements requires that $g(x_i)$ are of the same sign. Employing the method described in Ref.~\cite{PhysRevResearch.6.013279}, we optimize the objective function with a sequential quadratic programming algorithm, obtaining the optical coupling sequence $\left\{ x_i/x_0,g\left( x_i \right)/g_0  \right\} $ in real space, as listed in Table \ref{table1} [see Appendix.~\ref{appendix2}]. When $\delta_w$ is large, the numerical results agree well with the analytical results. However, when the width of the window function is narrow, the required number of coupling points $N$ in the optimization process will be large, which is very demanding experimentally. Therefore, we set $|k_r|/k_0=1.1$, and the coupling point number is $N=50$ in this work.

Furthermore, the optimal coupling function in momentum space is obtained by Fourier transformation $g_k=\sum_{x_i}{g\left( x_i \right) e^{-ikx_i}}$, which is illustrated in Fig.~\ref{fig3}. For the modes far away from the atomic transition frequency $\omega_0$, the coupling strength $g_k$ remains almost constant with respect to $k$. Due to $g(x_i)$ having the same sign, the coupling around $k\approx 0$ ( the dc component) is strong. Since the coupling strength is much smaller than $\omega_0$, the interaction between the system and low-frequency modes is weak, leading to a negligible contribution from modes around $c|k|=0$.

\subsection{Correlated photon pairs}
Based on the theoretical analysis in Sec.~\ref{section1_a}, when adopting the coupling function depicted in Fig.~\ref{fig3}, the correlated radiation of two photons can be achieved. To explore this two-photon process, we plot the evolution of the three-level giant atom shown in Fig.~\ref{fig4}(a). We find that the probability amplitudes of the giant atom being in the state $|e\rangle $ is extremely low, indicating that the $|e\rangle $ is only an unstable state during the correlated decay process. In Fig.~\ref{fig4}(b), the population in the unstable state decreases as $\omega_{eg}$ moves further from $\omega_0$. This reduction in state $ |e\rangle$ population contributes to the improved agreement between numerical and analytical results with increasing $\omega_{eg}$. The evolution of the state $|f\rangle $ approximately follows the form of exponential decay, demonstrating that the three-level atom decays from the state $|f\rangle $ to the ground state $|g\rangle $ by simultaneously releasing two photons into the waveguide. Therefore, the two photons are strongly correlated in space, and their second-order correlation function exhibits strong bunching.

We explore the wavefunction of two photons in both real and momentum space, focusing on the direction of their propagation. The atomic wavefunction, which describes the momentum distribution of two photons in $k$ space, is written as
\begin{gather}
	|\psi (k_1,k_2)\rangle=a_{k_1}^{\dagger}a_{k_2}^{\dagger}|\psi \left( t=0 \right) \rangle  ,
\end{gather}
where $a_{k_1(k_2)}^{\dagger}$ is the creation operator that creates a photon which wave vector is $k_1$ ($k_2$), with the initial state being $|f,0\rangle$. According to the numerical method in the Appendix.~\ref{appendix1}, we derive the wavefunction in $k$-space. 

The real space wavefunction for two correlated photons $|\psi \left( x_1,x_2 \right) \rangle$, is obtained via Fourier transform, which represents the probability of detecting two photons simultaneously at position $(x_1,x_2)$, that is
\begin{gather}
	|\psi \left( x_1,x_2 \right) \rangle =\sum_{k_1}{\sum_{k_2}{e^{ik_1x_1}e^{ik_2x_2}|\psi \left( k_1,k_2 \right) \rangle}}.
\end{gather}

The distribution of the two photons in real space is shown in Fig.~\ref{fig5}(a), with significant distributions in all four quadrants and the high intensity at $x=c\Delta t$, indicating the bidirectional propagation of the photons and their strong spatial correlation. Figure.~\ref{fig5}(b) shows the distribution in momentum space, demonstrating momentum conservation. The intensity is highest at $k_1=k_2=k_0$, which indicats that the three-level atom is most likely to simultaneously emit two photons with the same frequency $\omega_0$. We find that the two-photon distribution confirms the momentum conservation relationship $c\left| k_1 \right|+c\left| k_2 \right|=\omega_{fg} $. We take $x_0=2\pi/k_0$ as the units of the space length.

The second-order correlation function is defined as
\begin{gather}
	g^{\left( 2 \right)}\left( x_1,x_2 \right) =\frac{\left< a_{x_1}^{\dagger}a_{x_2}^{\dagger}a_{x_2}a_{x_1} \right>}{\left< a_{x_1}^{\dagger}a_{x_1} \right> \left< a_{x_2}^{\dagger}a_{x_2} \right>},
\end{gather}
where $a_{x_1(x_2)}^{\dagger}$ is the operator that creates a photon in the real space at position $x_1(x_2)$, with $r=x_1-x_2$ signifies the spatial separation between two photons, which are plotted in Fig.~\ref{fig5}(c) for the field evolution at $\Delta t=90$. Notably, the strongest correlation peaks occur both at $r=2c\Delta t$ and $r=0$,  correlation of the photons in the spontaneous emission process. In conclusion, in terms of spatial correlation, our scheme can realize bidirectional propagation of correlated two-photons. To ensure the validity of the Markov approximation, the propagation time between the first and the last coupling points of the giant atom must be much shorter than the spontaneous decay time, i.e., $(x_{N}-x_1)/c \ll  1/\Gamma$. In our work, the ratio $(x_{N}-x_1)\Gamma/c $ is approximately 0.1 [see Table.~\ref{table1}], which satisfies the Markov approximation.
 
\begin{figure}
	\centering\includegraphics[width=5cm]{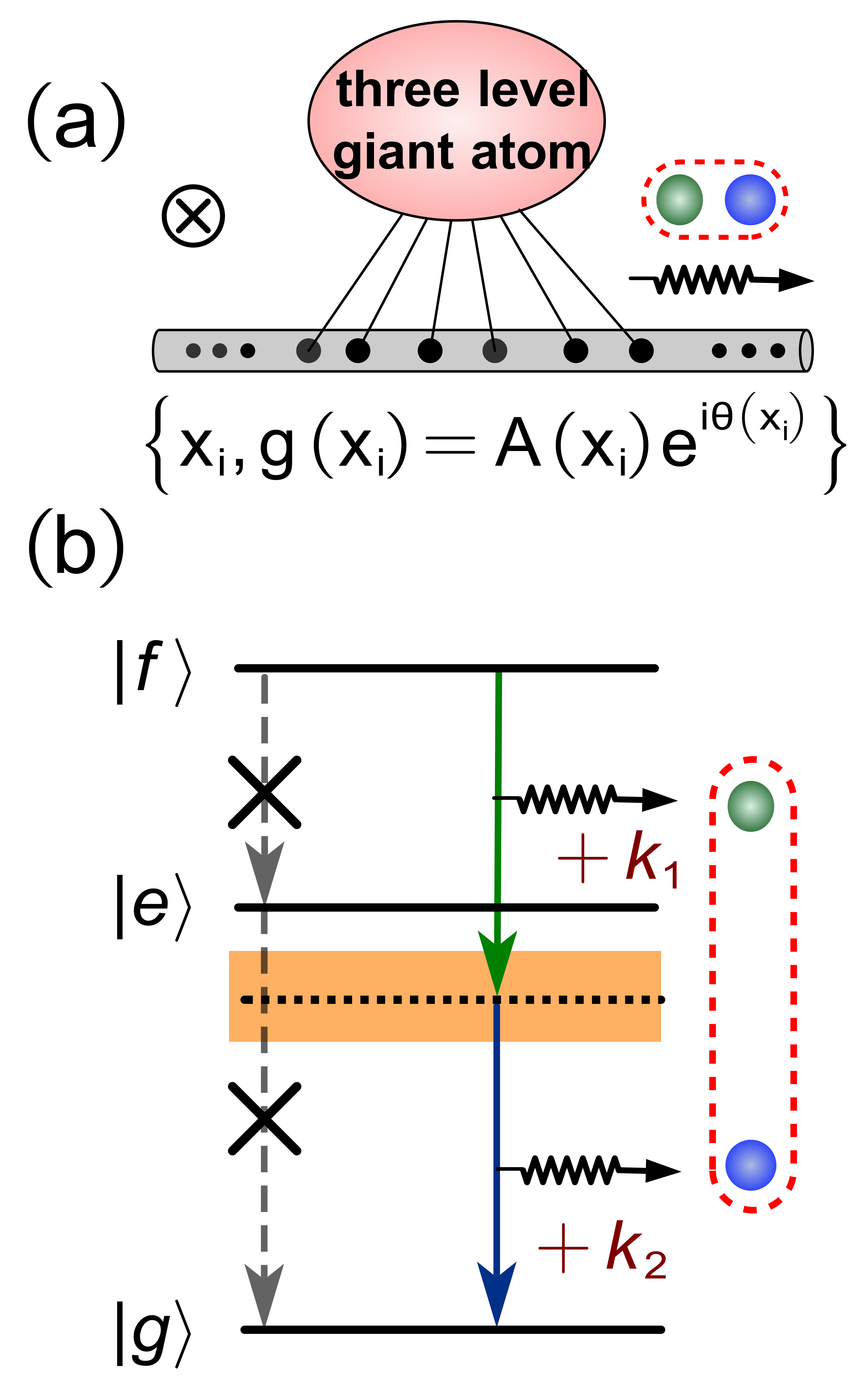}
	\caption{(a) Sketch of the chiral model: A three-level giant atom couples to the linear conventional wavguide, which photons can only travel in the right direction. (b) Energy level diagram of the ladder-type three-level giant atom.}
	\label{fig6}
\end{figure}
\begin{figure}
	\centering\includegraphics[width=8cm]{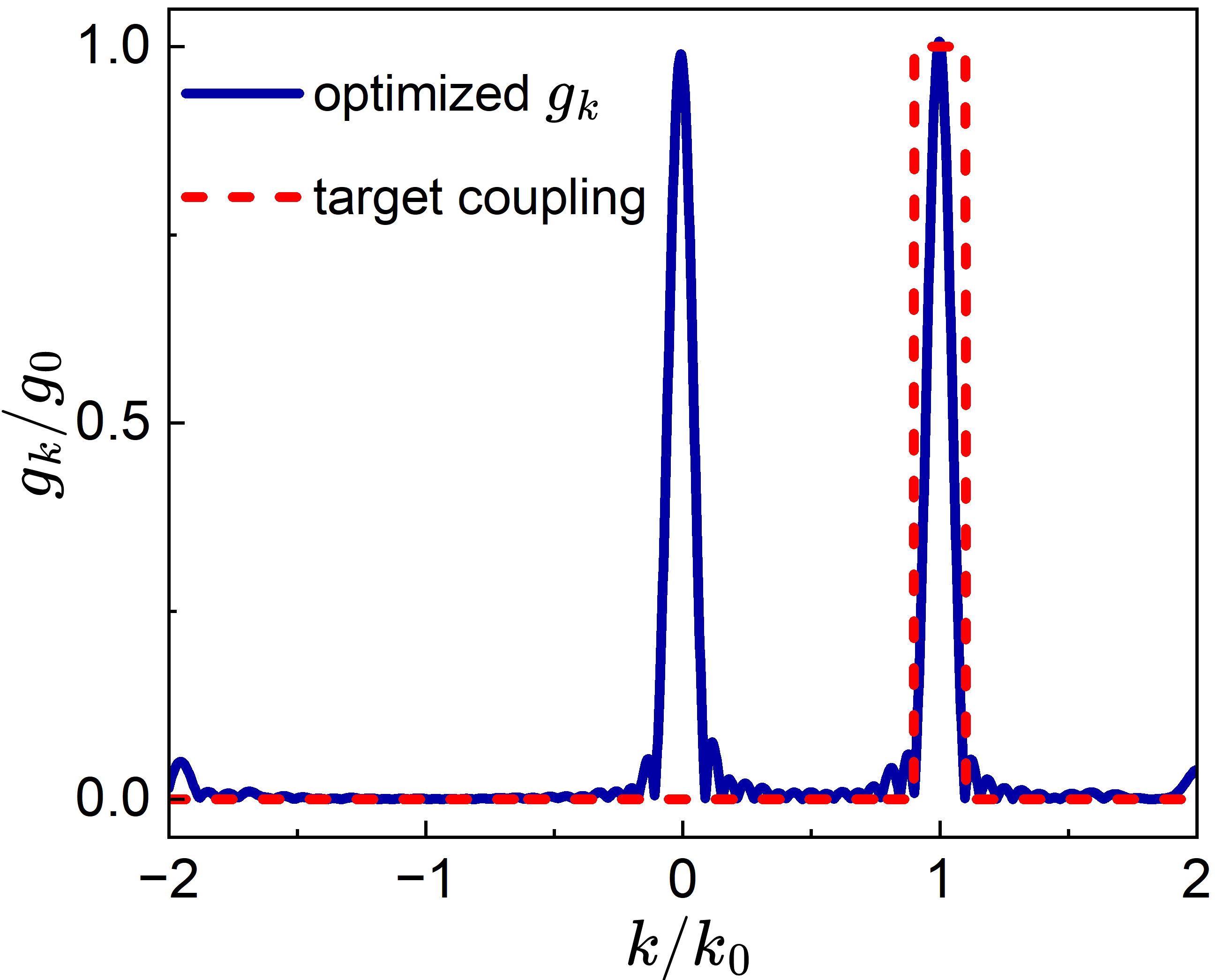}
	\caption{The optimizing coupling sequence in k-space for chiral case. The dashed blue line represents the objective function, and the solid orange line is the optimized coupling function. The real-space coupling sequence $\left\{ x_i/x_0,A\left( x_i \right)/g_0 ,\theta \left( x_i \right) \right\} $ is listed in Table.~\ref{table2} [see Appendix.~\ref{appendix2}]. The blue region represents the discrepancy between the objective function and the optimized result.}
	\label{fig7}
\end{figure}
\begin{figure*}
	\centering\includegraphics[width=17cm]{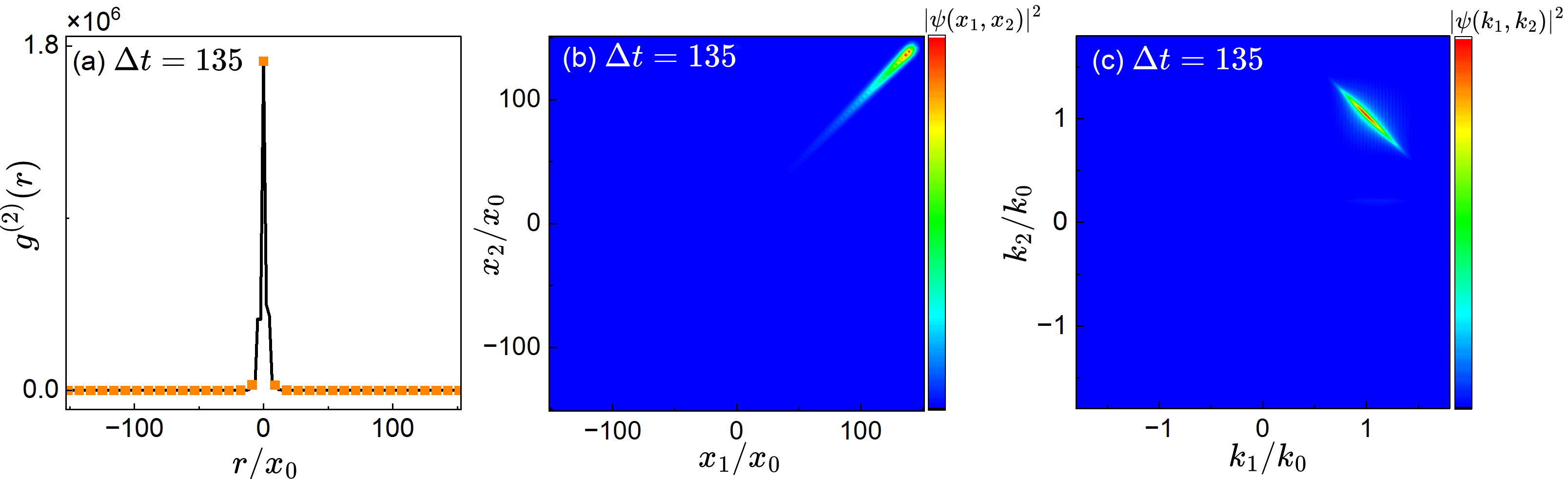}
	\caption{(a) The second-order correlation function of two photons in real space at $\Delta t=135$. (b) The two-photon field distribution in real space. (c) The two-photon field distribution in momentum space. Other parameters are the same with those in Fig.~\ref{fig5}.}
	\label{fig8}
\end{figure*}
\section{Directional two-photon chiral emission}
In our bidirectional propagation case study, we assume that the giant atom interacts with the waveguide via linear conventional elements. Therefore, there is no phase difference between the coupling strength of each coupling point. However, if we consider coupling with tunalbe couplers, such as Josephson junctions, a local phase $\theta(x_i)$ is encoded at $x_i$ by applying time-dependent magnetic flux through the loop, then the coupling strength is $g\left( x_i \right) =A\left( x_i \right) e^{i\theta \left( x_i \right)}$ \cite{PhysRevLett.113.220502,RevModPhys.93.025005}. The introduced phase factor affects the coupling between the emitter and different momentum modes, resulting in an asymmetry where the coupling strength satisfy $|g_k|\ne |g_{-k}|$ in momentum space, enabling directional control of photon emission.
	
Moreover, it has been experimentally shown that the Markov approximation is valid in such systems, supporting the applicability of our model. Additionally, within a specific range of detuning, the Wigner-Weisskopf theory remains valid, ensuring the accuracy of our theoretical framework under these conditions \cite{PhysRevX.13.021039,Wang_2022}.

The objective function in the form of window function is designed as
\begin{gather}
	g_{k}^{T}=\begin{cases}
		1,\quad \left\{ \omega_0-\delta_w <c\left| k \right|< \omega_0+\delta_w \right\} ,\\
		0,\quad \mathrm{others} .
	\end{cases}
	\label{g_k01}
\end{gather}

According to the same coupling method discussed in Sec.~\ref{sectionb}, we obtain the optimized coupling sequence $\left\{ x_i/x_0,A\left( x_i \right)/g_0 ,\theta \left( x_i \right) \right\} $ in real space with $N=50$. The corresponding data of the real space coupling sequence under the chiral case are presented in Table \ref{table2} [see Appendix.~\ref{appendix2}].

By employing the Fourier transform, we illustrate the optimized coupling function in momentum space in Fig.~\ref{fig7}. For modes around $k_0$, the coupling strength $g_k$ varies slightly with $k$, but it approaches zero at $k_{fe}$ ($k_{eg}$) and $k_0$ in the left direction. Similarly, the effect of the strong coupling between the emitter and the waveguide at $k\approx 0$ on our dynamics can be considered negligible. 

Under these conditions, the giant atom only interacts with the modes which propagates to the right direction, enabling the observation of correlated two-photon emission, as shown in Fig.~\ref{fig6}(a). Initially, a virtual process occurs where the giant emitter transitions from the $|f\rangle$ state to the intermediate state $|m\rangle$, releasing a photon with momentum around $ +k_1$ which propagates to the right direction. Due to the non-conservation of energy, the atom keeps on rapidly decaying from the intermediate state $|m\rangle $ to the ground state $|g\rangle $ by emitting a photon $+k_2$, as shown in Fig.~\ref{fig6}(b). We find that these two photons are released simultaneously and propagate to the same direction, leading to a strong spatial correlation in one direction.

As depicted in Fig.~\ref{fig8}(a), the two-photon correlation peaks exclusively at $r=0$, indicating that the two photons only propagate in the same direction. To further investigate the propagation direction of the photons, we focus on the two-photon distributions in real space in Fig.~\ref{fig8}(b). The results reveals a distribution predominantly in the right direction,  thus confirming the right emission. Furthermore, in momentum space, we observe the satisfaction of energy conservation $ ck_1 + ck_2 =\omega_{fg} $, which is depicted in Fig.~\ref{fig8}(c). Therefore, we can achieve directional correlated two-photon emission without energy loss. The same approach can be taken for photon pairs that achieve propagation in the left direction.

\section{Two-photon mediated cascaded quantum system}

\begin{figure}[h]
	\centering\includegraphics[width=8.6cm]{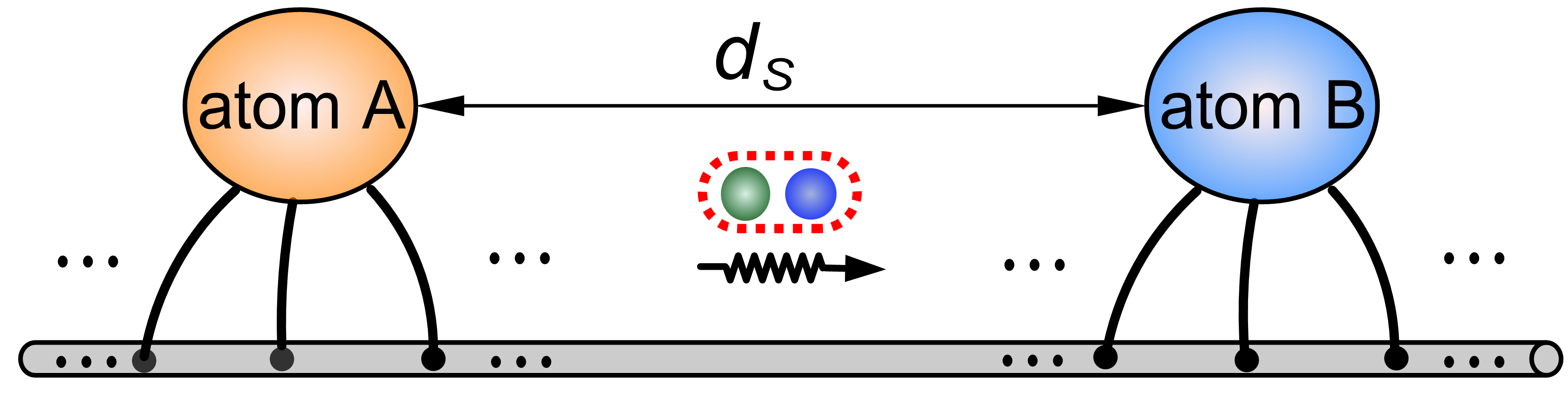}
	\caption{Diagram of a two-photon mediated cascaded quantum system, illustrating the emission and rightward propagation of two photons between two distant three-level atoms. }
	\label{fig9}
\end{figure}
In this section, based on the directional two-photon chiral emission 
discussed above, we consider the two-photon mediated cascade quantum network 
for three-level giant atoms. We take the simplest setups where two giant 
atoms 
are coupled to a conventional linear waveguide for example (cf. 
Fig.~\ref{fig9}). The coupling centers of two giant atoms are  
$x_{A,B}$.  Now we derive the cascaded master equation for 
this two-photon mediated system. Given that two giant atoms are identical and 
two 
photons are emitted simultaneously, the interaction Hamiltonian for 
Eq.~(\ref{gkk}, \ref{c_f1}) is given by
\begin{eqnarray}
	H_{\mathrm{int}}&=&\sum_{i=A,B}{\sum_{k,k'}{g_{i,kk'}e^{i\delta_{k,k'} 
	t}\sigma _{i}^{+}a_ka_{k'}+\mathrm{H}.\mathrm{c}.}},
	\label{Hint1}
	\\
	g_{i,kk'}&=&\frac{g_{ik}g_{ik'}}{\omega _k-\omega _{eg}},\quad 
	g_{Bk}=g_{Ak}e^{ikd_s},
	\label{H_int}
\end{eqnarray}
where $g_{k,k'}$ represents the two-photon coupling strength, and $\sigma 
^+=|f\rangle \langle g|$ denotes the atom transition operator. Here, $d_s$ is the separation distance between two giant atoms. Due to the 
state $|e\rangle$ being adiabatically eliminated, the giant atoms are 
treated as two-level qubits which decay to their ground state by emitting two 
correlated photons. In this scenario, two giant atoms are unidirectionally 
mediated by
correlated photon pairs. The cascaded master equation governing the system's 
evolution can be derived as \cite{Wang_2022,scully_zubairy_1997}
\begin{equation}
	\begin{split}
		\dot{\rho}_s\left( t \right) &=-i\mathrm{Tr_R}\left[ H_{\mathrm{int}},\rho \left( t \right) \right]  \\
		&-\mathrm{Tr_R}\int_{t_0}^t{\left[ H_{\mathrm{int}}\left( t \right) ,\left[ H_{\mathrm{int}}\left( t' \right) ,\rho \left( t' \right) \right] \right] dt'},
	\end{split}
	\label{master}
\end{equation}
with $\rho$ ($\rho_s$) being the density matrix operator of the system (two 
emitters). By substitution Eq.~(\ref{H_int}) into Eq.~(\ref{master}), we 
obtain \cite{PhysRevA.99.053852}
\begin{eqnarray}
	\dot{\rho}_s\left( t \right) 
	=&&-\sum_{i,j}{\int_{t_0}^t{dt'A\left(x_i,x_j;t,t' 
	\right)}}\notag\\
	&&\left( \sigma _{i}^{+}\sigma _{j}^{-}\rho _s\left( t' \right) -\sigma 
	_{j}^{-}\rho _s\left( t' \right) \sigma _{i}^{+} \right) 
	+\mathrm{H}.\mathrm{c}.,
	\label{rho_t1}
\end{eqnarray}
with $A\left(x_i,x_j;t,t' 
\right)$ being the spatiotemporal correlation function 
for the propagating field. According to Eq.~(\ref{c_f2}), it is defined as
\begin{eqnarray}
		A\left(x_i,x_j;t,t' 
		\right) 
		={\sum_{k,k'}{g_{i,kk'}g_{j,kk'}^{\ast}e^{i\delta_{k,k'}
		 \left( t-t' \right)}}} \notag \\
		\approx \frac{L^2|g_{k_0}|^4}{4\left( 2\pi \right) ^2c^{2}}\frac{\delta 
		_w}{\Delta ^2-\delta _{w}^{2}}2\pi \delta \left( 
		t-t'-\frac{x_{ij}}{c} 
		\right),
	\label{A_xt}  
\end{eqnarray}
 where  $x_{ij}=x_i-x_j$.
 In our proposal, we consider the waveguide to be a linear medium 
 $\omega_k=c|k|$.
  When the separation distance $d_s$ is significantly smaller 
 than the size of wavepacket, the time-retardation effect is negligible. 
 Given that two atoms
 chirally decay/absorb two correlated photons,  
 the master equation is simplified as~\cite{Wang_2022}
\begin{eqnarray}
		\dot{\rho}_s\left( t \right) &=&\frac{\Gamma 
		_R}{2}\sum_{i=A,B}{\left[ 
		\sigma _{i}^{+}\sigma _{i}^{-}\rho_s(t) -\sigma _{i}^{-}\rho _s(t)\sigma 
		_{i}^{+} \right]} \notag \\
		&&-\Gamma _R\left[ \sigma _{B}^{+}\sigma _{A}^{-}\rho _s\left( t 
		\right) -\sigma _{A}^{-}\rho _s(t)\sigma _{B}^{+} 
		\right]+\mathrm{H}.\mathrm{c}., 
\end{eqnarray}
where $$\Gamma_R=\frac{2L^2|g_{k_0}|^4}{\pi c^2}\frac{\delta _w}{\Delta ^2-\delta _{w}^{2}},$$ is the rate of the spontaneous decay to the right 
	direction. The evolution of the system can be described by
\begin{gather}
	\dot{\rho}_s\left( t \right) =-iH_{\mathrm{eff}}\rho _s+i\rho _sH_{\mathrm{eff}}^{\dagger}+\mathcal{L} \rho _s\mathcal{L} ^{\dagger},
	\notag\\
	H_{\mathrm{eff}}=-i\sum_{i=A,B}{\frac{\Gamma _R}{2}\sigma _{i}^{+}\sigma 
	_{i}^{-}-i\Gamma _R}\sigma _{B}^{+}\sigma _{A}^{-},
	\label{Heff}
\end{gather}
with $ \mathcal{L}$ being the collective jump operator, $\mathcal{L} 
=\sqrt{\Gamma _R}\left( \sigma _{A}^{-}+\sigma _{B}^{-} \right) $. From the 
last term of the Eq.~({\ref{Heff}}), we derive that the two-photon state 
emitted by emitter A is reabsorbed by the emitter B, but does not flow back 
into the emitter A. Note that the cascaded master equation ignores the time 
decay effects by tracing over the waveguide's degrees of freedom. To overcome 
these constraints, we employ a simulation based on the unitary dynamics 
governed by the Hamiltonian Eq.~(\ref{Hint1}). As discussed in Appendix.~\ref{appendix1}, 
this numerical method  can effectively capture the time delay effects.
\begin{figure*}
	\centering\includegraphics[width=17.5cm]{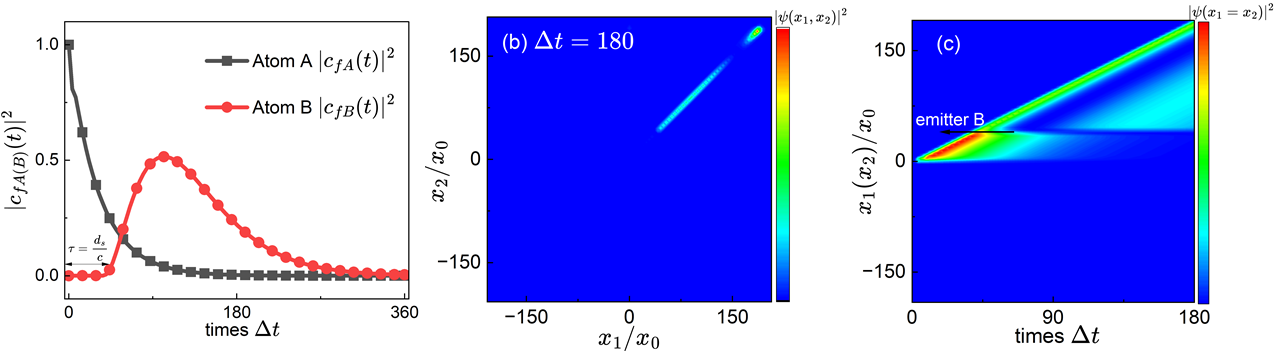}
	\caption{(a) The evolution of probability amplitudes between two giant 
	emitters. Initially, emitter A is excited and spontaneously decay. After 
	a retardation time, emitter B is excited at $t=d_s/c$. (b) The two-photon distribution in real space at $\Delta t=180$. (c) The 
	distribution of the field along the $x_1=x_2$ direction varies with time. We 
	set $d_s/x_0=40$ and $g=0.018$.}
	\label{fig10}
\end{figure*}

\begin{figure}[h!]
	\centering\includegraphics[width=7.5cm]{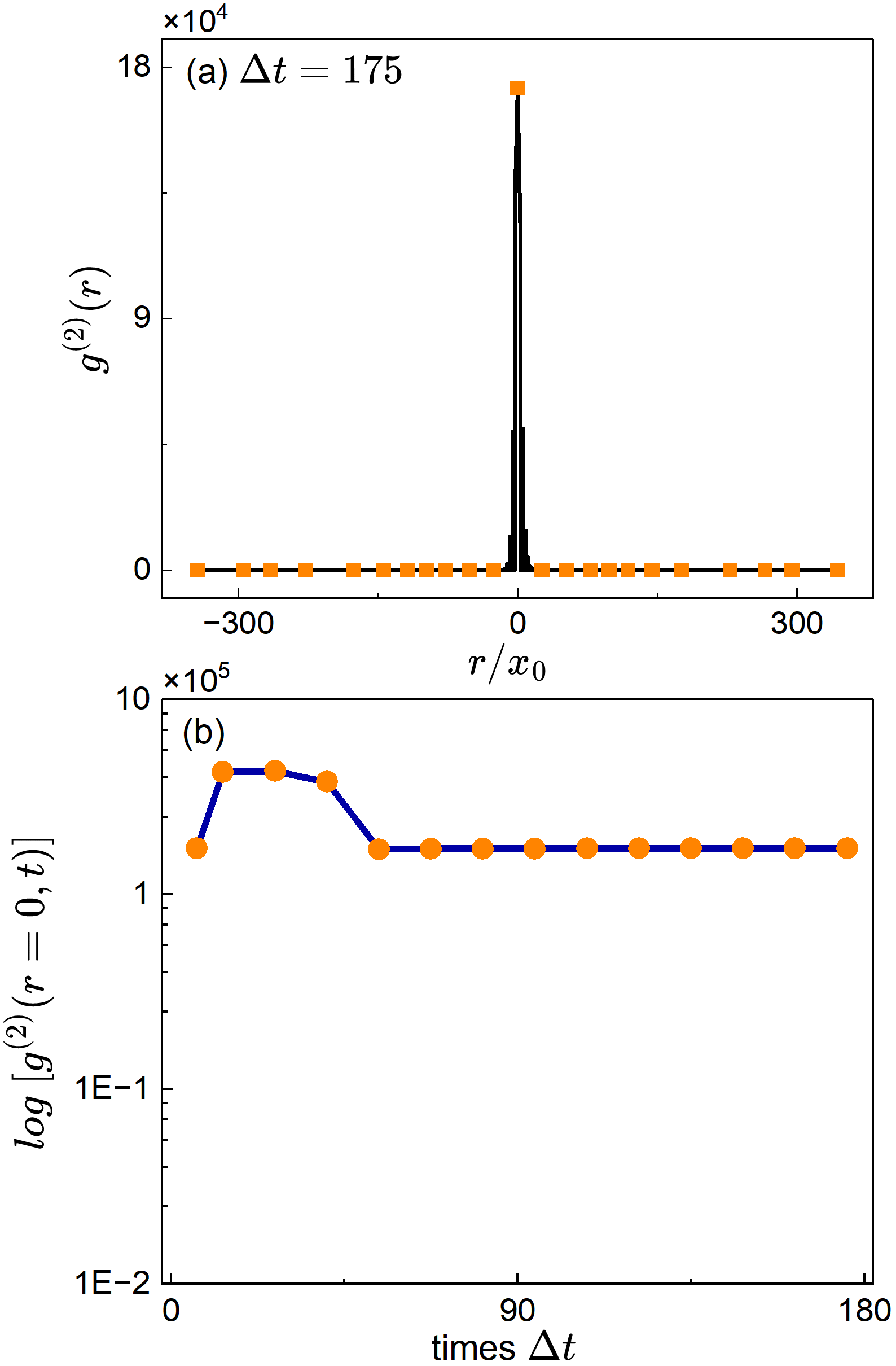}
	\caption{(a) The second-order correlation of two photons at $\Delta t=175$. (b) The second-order correlation function at $r=0$ varies with time. The parameters are the same with those in Fig.~\ref{fig10}.}
	\label{fig11}
\end{figure}

 Assuming that the emitter A is initially in $|f\rangle$, 
 Fig.~\ref{fig10}(a) 
 illustrates the temporal evolution of the population on the two emitters 
 being in $|f\rangle$ (denoted as $|c_{fA(B)(t)}|^2$). Based on Fig.~\ref{fig10}(b), we observe that two strongly correlated photons are generated in the system.  As depicted in Fig.~\ref{fig11}(b), we find that, due to the absorption and re-emission by the emitter B, the correlated two photons will be scattered slightly. However, they remain strongly bunched ($g^{(2)}(r=0,t)\approx10^5$), preserving their high degree of correlation. It is clear that emitter B reabsorbs the correlated photon pair from emitter A. In parallel, as depicted in Fig.~\ref{fig10}(c), we show the evolution of the field distribution in real space along the $x_1=x_2$ direction over time. We find that at the initial of the evolution, the emitter A decays spontaneously, emitting a two-photon field that propagates in the right direction. After a propagating time $t=d_s/{c}$, the emitted field reaches emitter B, and be absorbed without being scattered back to emitter A, indicating that the two-photon mediated cascaded system is realized. Compared with chiral setups mediated by single photon, our proposal can mediate three-level giant atoms, and two correlated photons 
 serve as correlated flying qubits, which might underpin unique application in many-body physics and quantum information processing~\cite{wang2024nonlinear,wang2024longrange}.

\section{Conclusion}
In this paper, we explore how to realize two-photon correlated emission by considering a three-level giant atom interacting with a conventional linear waveguide. In contrast to previous studies, this proposal does not need nonlinear medium. By optimizing the objective coupling function using sequential quadratic programming algorithm, the emission of a single photon is prevented, while the process of two correlated photons being emitted from the giant atom is significantly enhanced. Meanwhile, the optimal coupling function in momentum space are illustrated by Fourier transform. Considering the emitter coupling to the waveguide via linear elements, the coupling strength in right and left directions are symmetric. We show the spatial field distribution and the second-order correlation function, revealing the bidirectional propagation of photons in the waveguide and the strong spatial correlations between two identical itinerant photons. Meanwhile, this scheme ensures energy conservation without photon loss, thereby addressing the issue of photon loss in multi-photon dynamics.

Similarly, when the coupling element are adopted as nonlinear element, local 
phase are encoded at each coupling point. When the giant atom interacting 
with the waveguide in either the right or left direction, we achieve 
directional chiral two-photon correlated emission, which both photons 
propagate in the same direction. Since our proposal does not need nonlinear 
medium, the efficiency of emitting correlated photons can be much higher. 
Moreover, we show that the emitted two photons can mediate remote three-level 
atoms, indicating our work can enable the cascaded process of quantum 
information in qutrit systems, which can be configured for quantum 
information and many-body quantum physics.
 
\section{Acknowledgments}
The quantum dynamical simulations are based on open source code 
QuTiP ~\cite{Johansson12qutip,Johansson13qutip}. 
X.W.~is supported by 
the National Natural Science
Foundation of China (NSFC) ( No.~12174303 and Grant No.~11804270), and the Fundamental 
Research Funds for the Central Universities (No. xzy012023053). W.X.L. is supported by the Natural Science Foundation of Henan Province (No. 222300420233).
\appendix
\section{The numerical simulation method}
\label{appendix1}
We briefly introduce the numerical method in this work. The specific steps are as follows:

(a) We discretize a set of $M=2000$ modes within the range of $k\in \left( -2k_0 ,2k_0 \right] $, corresponding to the discretization of a finite waveguide with a length $L=2000\lambda_0 $ in real space, with $\lambda_0=2\pi/k_0$. The considerable length of $L$ is crucial in preventing the propagating wavepacket from reaching the simulation boundary of the waveguide.

(b) In the two-excitation subspace, the state of the system is denoted as
$
|\psi \left( t \right) \rangle =c_f\left( t \right) |f,0\rangle +\sum_k{c_{e,k}\left( t \right)|e,k\rangle}+\sum_{k,k\prime}{c_{g,k,k\prime}\left( t \right) |g,k,k'\rangle}   .
$
The dimensions of the Hamiltonian matrix is $M^2+M+1$, which is expressed as 
\begin{widetext}
	\begin{eqnarray*}
		\left( \begin{smallmatrix}
			\omega _f&		g_{k1}&		g_{k2}&		\cdots&		g_{kM}&		0&		0&		\cdots&		0&		0&		\cdots&		0&		\cdots&		0&		\cdots&		0\\
			g_{k1}^{\ast}&		\omega _{k1}+\omega _{eg}&		0&		\cdots&		0&		g_{k1}&		g_{k2}&		\cdots&		g_{kM}&		0&		\cdots&		0&		\cdots&		0&		\cdots&		0\\
			g_{k2}^{\ast}&		0&		\omega _{k2}+\omega _{eg}&		\ddots&		0&		0&		0&		\cdots&		0&		g_{k1}&		\cdots&		g_{kM}&		\cdots&		0&		\cdots&		0\\
			\vdots&		\vdots&		\ddots&		\ddots&		0&		0&		0&		\cdots&		0&		0&		\cdots&		0&		\cdots&		0&		\cdots&		0\\
			g_{kM}^{\ast}&		0&		0&		0&		\omega _{kM}+\omega _{eg}&		0&		0&		\cdots&		0&		0&		\cdots&		0&		\cdots&		g_{k1}&		\cdots&		g_{kM}\\
			0&		g_{k1}^{\ast}&		0&		0&		0&		\omega _{k1}+\omega _{k1}&		0&		\cdots&		0&		0&		\cdots&		0&		\cdots&		0&		\cdots&		0\\
			0&		g_{k2}^{\ast}&		0&		0&		0&		0&		\omega _{k1}+\omega _{k2}&		\ddots&		0&		0&		\cdots&		0&		\cdots&		0&		\cdots&		0\\
			\vdots&		\vdots&		\vdots&		\vdots&		\vdots&		\vdots&		\vdots&		\ddots&		0&		\ddots&		\cdots&		0&		\cdots&		0&		\cdots&		0\\
			0&		g_{kM}^{\ast}&		0&		0&		0&		0&		0&		0&		\omega _{k1}+\omega _{kM}&		0&		\cdots&		0&		\cdots&		0&		\cdots&		0\\
			0&		0&		g_{k1}^{\ast}&		0&		0&		0&		0&		0&		0&		\omega _{k2}+\omega _{k1}&		0&		0&		\cdots&		0&		\cdots&		0\\
			\vdots&		\vdots&		\vdots&		\vdots&		\vdots&		\vdots&		\vdots&		\vdots&		\vdots&		\ddots&		\ddots&		0&		\cdots&		0&		\cdots&		0\\
			0&		0&		g_{kM}^{\ast}&		0&		0&		0&		0&		0&		0&		0&		0&		\omega _{k2}+\omega _{kM}&		0&		0&		\cdots&		0\\
			\vdots&		\vdots&		\vdots&		\vdots&		\vdots&		\vdots&		\vdots&		\vdots&		\vdots&		\vdots&		\vdots&		\ddots&		\ddots&		\ddots&		\ddots&		0\\
			0&		0&		0&		0&		g_{k1}^{\ast}&		0&		0&		0&		0&		0&		0&		0&		0&		\omega _{kM}+\omega _{k1}&		0&		0\\
			\vdots&		\vdots&		\vdots&		\vdots&		\vdots&		\vdots&		\vdots&		\vdots&		\vdots&		\vdots&		\vdots&		\vdots&		\ddots&		\ddots&		\ddots&		0\\
			0&		0&		0&		0&		g_{kM}^{\ast}&		0&		0&		0&		0&		0&		0&		0&		0&		0&		0&		\omega _{kM}+\omega _{kM}\\
		\end{smallmatrix} \right) 
	\end{eqnarray*}
\end{widetext}
where $ \omega_{k_i} $ is the frequency at mode $k_i$ and  $ g_{k_i} $ is the coupling strength between the giant atom with mode $ k_i $. 

(c) We assume the giant atom initially in the second excited state and these is no photon in the waveguide, i.e., $ |\psi \left( t=0 \right) \rangle = |f,0\rangle $, we employ Qutip package ~\cite{Johansson12qutip,Johansson13qutip} to numerically solve the Schr$ \ddot{\mathrm{o}} $dinger equation. We can obtain the probability of the three-level giant emitter and the field distribution in both real space and momentum space. 

\section{Optimized coupling sequence}
\label{appendix2}
\begin{table*}
	\renewcommand\arraystretch{1.4}
	\begin{tabular}{>{\hfil}p{1in}<{\hfil}>{\hfil}p{0.53in}<{\hfil}>{\hfil}p{0.53in}<{\hfil}
			>{\hfil}p{0.53in}<{\hfil}>{\hfil}p{0.53in}<{\hfil}>{\hfil}p{0.53in}<{\hfil}>{\hfil}p{0.53in}<{\hfil}
			>{\hfil}p{0.53in}<{\hfil}>{\hfil}p{0.53in}<{\hfil}>{\hfil}p{0.53in}<{\hfil}>{\hfil}p{0.53in}<{\hfil}}
		\hline \hline  
		\text{$x_i/x_0$}& -6.97974&	-6.1012&  -5.85101&  -5.13147 &  -4.87762&   -4.62711&	-4.37708&  -4.12705& -3.87654& -3.6265\\
		\hline 
		\text{$g(x_i)/g_0$}&0.01244&	0.024889&	0.019228&	0.035245&	0.036471&	0&	0&	0.04685&	0.047951 &0\\
		\hline  \hline
		\text{$x_i/x_0$}&	-3.37631&	-3.12612&  -2.87609&	-2.62542&	-2.37539&	-2.12536& -1.87532&	-1.62497& -1.37492&   -1.12481\\
		\hline 
		\text{$g(x_i)/g_0$}& 0&	0.056732&	0.057731&	0&  0&	0.064517&	0.064959&	0&   0&	  0.069555 \\
		\hline  \hline
		\text{$x_i/x_0$}&	-0.87483&	-0.62465&	-0.37443&	-0.12434&	0.125658&	0.376449& 0.626466&	0.876546&1.126562&	1.376579 \\
		\hline 
		\text{$g(x_i)/g_0$}&	0.069423&	0&	0&	0.0717&	0.070923&	0&	0&	0.070929 & 0.06927 & 0 \\
		\hline \hline 
		\text{$x_i/x_0$}&	1.626723&	1.877073&	2.127106&	2.377138&	2.62733&	2.878317&	3.12835&	3.381033& 3.630324&	3.880516 \\
		\hline 
		\text{$g(x_i)/g_0$}&	0&	0.067023&	0.064766&	0&	0&	0.060415&	0.057325&	0 & 0 & 0.051466 \\
		\hline\hline
		\text{$x_i/x_0$}&	4.130548&	4.381695&	4.885579&	5.140068&	5.628992&	5.882367&	6.141789&	6.40344& 6.881541&	7.13237\\
		\hline 
		\text{$g(x_i)/g_0$}&	0.047489&	0&	0.041289&	0.035751&	0&	0.027585&	0.02458&	0& 0.012783 & 0.011244 \\
		\hline\hline
	\end{tabular}
	\caption{The optimized coupling sequence in the case of bidirectional photon propagation in real space.}
	\label{table1}
\end{table*}
\begin{table*}
	\renewcommand\arraystretch{1.4}
	\begin{tabular}{>{\hfil}p{1in}<{\hfil}>{\hfil}p{0.53in}<{\hfil}>{\hfil}p{0.53in}<{\hfil}
			>{\hfil}p{0.53in}<{\hfil}>{\hfil}p{0.53in}<{\hfil}>{\hfil}p{0.53in}<{\hfil}>{\hfil}p{0.53in}<{\hfil}
			>{\hfil}p{0.53in}<{\hfil}>{\hfil}p{0.53in}<{\hfil}>{\hfil}p{0.53in}<{\hfil}>{\hfil}p{0.53in}<{\hfil}}
		\hline \hline  
		\text{$x_i/x_0$}& -7.14176&	-6.86085&  -6.57867&  -6.27723 &  -5.99664&   -5.71971&	-5.44167&  -5.16728& -4.88574& -4.59289\\
		\hline 
		\text{$A(x_i)/g_0$}&0.005645&	0.009507&	0.00452&	0.0083&	0.018594&	0.015474&	0.000308&	0.021212&	0.030514 &0.013119\\
		\hline
		\text{$\theta(x_i)$}& -0.30258&	0.70513&  1.5708&  -0.65417 &  0.27106&   1.1329&	1.5708&  -0.34866& 0.62847& 1.4027\\
		\hline  \hline
		\text{$x_i/x_0$}&	-4.22891&	-3.94545&  -3.63653&	-3.35117&	-3.06548&	-2.76914& -2.48218&	-2.19666& -1.90779&   -1.61892\\
		\hline 
		\text{$A(x_i)/g_0$}& 0.022308&	0.040937&	0.025162&	0.014754&  0.045659&	0.040736&	0.002907&	0.040073&   0.05328&	  0.024942 \\
		\hline
		\text{$\theta(x_i)$}& -0.5722&	0.38919&  1.2498&  -1.0289 &  -0.00959&   0.847&	1.5708&  -0.46378& 0.42104& 1.3122\\
		\hline  \hline
		\text{$x_i/x_0$}&	-1.33496&	-1.05367&	-0.77144&	-0.48584&	-0.19703&	0.077228& 0.35786&	0.642509&0.92504&	1.21284 \\
		\hline 
		\text{$A(x_i)/g_0$}&	0.023473&	0.055365&	0.045693&	0.000698&	0.044941&	0.05631&	0.027011&	0.022645 & 0.056513 & 0.046022 \\
		\hline
		\text{$\theta(x_i)$}& -0.97905&	-0.07208&  0.82751&  1.5708 &  -0.59173&   0.32023&	1.1815&  -1.1586& -0.21906& 0.70155\\
		\hline \hline 
		\text{$x_i/x_0$}&	1.801634&	2.081588&	2.370613&	2.67635&	2.956144&	3.259493&	3.558705&	3.850913& 4.158241&	4.446152 \\
		\hline 
		\text{$A(x_i)/g_0$}&	0.044359&	0.051272&	0.019786&	0.025138&	0.049259&	0.032616&	0.009067&	0.039377 & 0.034449 & 0.003942 \\
		\hline
		\text{$\theta(x_i)$}& -0.67889&	0.27091&  1.0651&  -1.1264 &  -0.16069&   0.72162&	-1.5708&  -0.51485& 0.39107& 1.5708\\
		\hline\hline
		\text{$x_i/x_0$}&	4.744091&	5.034389&	5.316889&	5.595251&	5.893508&	6.17362&	6.455643&	6.732413& 7.009979&	7.322878\\
		\hline 
		\text{$A(x_i)/g_0$}&	0.024155&	0.029487&	0.013626&	0.008667&	0.020083&	0.014884&	0.001918&	0.007753& 0.009316 & 0.004281 \\
		\hline
		\text{$\theta(x_i)$}& -0.9162&	-0.01658&  0.95508&  -1.4322 &  -0.51606&   0.44446&	1.4502&  -1.0517& -0.08926& 0.76751\\
		\hline\hline
	\end{tabular}
	\caption{The optimized coupling sequence in the case of directional photon propagation in real space.}
	\label{table2}
\end{table*}


%

\end{document}